\documentclass[a4paper,12pt]{article}
\usepackage{amsmath}
\usepackage{amssymb}
\usepackage{newlfont}
\usepackage{stmaryrd}
\usepackage{mathrsfs}
\usepackage{mathtools}
\usepackage{euscript}
\usepackage{siunitx}
\usepackage{authblk}
\newcommand{\id}{\textrm{d}}

\let\oldsqrt\sqrt
\def\sqrt{\mathpalette\DHLhksqrt}
\def\DHLhksqrt#1#2{%
	\setbox0=\hbox{$#1\oldsqrt{#2\,}$}\dimen0=\ht0
	\advance\dimen0-0.2\ht0
	\setbox2=\hbox{\vrule height\ht0 depth -\dimen0}%
	{\box0\lower0.4pt\box2}}

\title{On the derivation of the Kompaneets equation}
\author{Guilherme Eduardo Freire Oliveira}
\author{Christian Maes}
\author{Kasper Meerts}
\affil{Instituut voor Theoretische Fysica, KU Leuven}
\date{}

\begin{document}
\maketitle
\begin{abstract}
The relaxation of a photon bath to thermal equilibrium via Compton scattering 
with electrons is described in the Kompaneets equation (1956). The equation is 
mostly known from studies of astrophysical plasmas, for its convergence to the 
Planck distribution and for possible corrections to that Planck law in the cosmic
microwave background, most notably from the Sunyaev-Zeldovich effect. We revisit its derivation 
emphasizing its structure as a Kramers-Moyal diffusion approximation to the 
quantum Boltzmann equation or Master equation with stimulated emission. We do 
not assume that the Planck law is stationary in performing the continuum 
approximation but we emphasize the necessity of the flux or M{\o}ller factor to 
arrive at a continuity equation. On the other hand, the structure allows more 
general assumptions than originally envisioned by Kompaneets.
\end{abstract}

\section{Introduction}
The interaction between matter and radiation is a standard topic in plasma 
physics and cosmology. In low-density plasmas, Compton 
scattering is the dominant mechanism enabling energy exchange \cite{blumenthalgould, zeldovich}.  Essentially, this process 
describes elastic collisions of photons and charged particles (which we assume to consist solely of 
electrons), preserving 
photon number.
The 
time evolution of the radiation spectrum subject to Compton scattering on a plasma 
of nonrelativistic Maxwellian electrons was first obtained by 
A.S.~Kompaneets\footnote{This arose from his theoretical research as part of 
the Soviet hydrogen bomb program in 1949. When the eponymous relaxation 
equation turned out to be useless for weapons research, the results were 
declassified and published by Kompaneets in 1956 \cite{peebles, 
longair}.} in 1956 \cite{kompa}. The Kompaneets equation later found new life in 
quantitatively describing the Sunyaev-Zeldovich effect, which is a distortion 
of the cosmic microwave background radiation by Compton scattering of hot 
electrons during its passage through clusters of galaxies \cite{sunyaeveffect, sunyaev, burigana3}.\\

For a nonrelativistic 
electron bath at temperature $T$, with $k_BT\ll 
m_ec^2$ for electron mass $m_e$, the Kompaneets equation
\begin{equation}\label{ke}
		\omega^2\frac{\partial n}{\partial t}(\omega,t)= \frac{n_e\sigma_T 
		c}{m_e c^2}\frac{\partial }{\partial \omega}\omega^4\left\{k_B T 
		\frac{\partial n}{\partial \omega}(\omega,t) + 
		\hbar\left[1+n(\omega,t)\right]n(\omega,t)\right\}
\end{equation}
shows how a photon gas under conditions of spatial homogeneity and isotropy 
relaxes to thermal equilibrium from Compton scattering. It is a continuity equation for the (dimensionless) photon 
number $n(\omega,t)$ occupying frequency $\omega$ at time $t$.   The spectral energy density 
is proportional to $\hbar(\omega/c)^3\,n(\omega,t)$ \cite{practical}.  The range of frequencies is assumed to have the photon energy $\hbar \omega \sim k_B T \ll m_ec^2$.  In \eqref{ke}, $\sigma_T \approx \SI{0.66}{\barn}$ is the total 
Thomson cross section and $n_e$ is the electron density, which together 
determine the photon mean free path $\ell = (n_e \sigma_T)^{-1}$ and the 
average time between collisions $\tau =\ell / c$.\\  
The current in frequency space is given by
\begin{eqnarray}\label{cur}
\frac{\partial n}{\partial t}(\omega,t)&=& \frac 1{\omega^2}\frac{\partial }{\partial \omega}\big(\omega^2\,j_t(\omega)\big)\nonumber\\
 j_t(\omega)&=& \frac{n_e \sigma_T c}{m_e c^2} \omega^2\left\{k_B T 
\frac{\partial n}{\partial \omega}(\omega,t) + 
\hbar\left[1+n(\omega,t)\right]n(\omega,t)\right\}
\end{eqnarray}
That current \eqref{cur} vanishes when the photon number $n(\omega,t)$ is given 
by the Bose-Einstein expression, \[ n_\text{eq}(\omega) = \frac{1}{\exp(\beta 
\hbar \omega) - 1}
\]
(with photon chemical potential taken zero) at inverse temperature $\beta = 
1/k_B T$. The temperature $T$ here refers to the temperature of the electron 
bath under the assumption that it is in thermal equilibrium, but this can be 
made more general as shown in Section \ref{gk}. The relaxation behavior of 
\eqref{ke} is obviously an interesting question in itself, but not the subject 
of the present paper; see e.g. \cite{burigana1,burigana2} for a numerical code.  
It suffices to note in addition that depending on the initial number of 
photons, the limiting photons as $t\uparrow\infty$ will have a Bose-Einstein 
distribution component ($n_\text{eq}(\omega)$), as well as a condensate (for 
$\omega = 0$). That condensate appears when the initial number of photons is 
larger than the equilibrium number of photons $n_\text{eq}$ at the given 
inverse temperature $\beta$. The excess photons cannot disappear, forming a 
Bose-Einstein condensate at $\omega=0$, as the Kompaneets equation is 
photon-number preserving. For this interesting feature we refer to 
\cite{escobedo1, levermore} for a more detailed discussion.\\ 

Kompaneets himself only derived the exact form of the first term between the 
brackets in \eqref{ke}, which is purely a diffusive term due to Doppler shift experienced by 
the photons in the reference frame of the electrons.
The (second) drift term, describing stimulated emission and the Compton recoil, were 
instead derived from the assumption that in thermal equilibrium the stationary 
distribution ought to be the Planck distribution. The so-called Comptonization, 
i.e., the redistribution of photon energies scattered by electrons consists of 
two terms indeed, both suppressed by the same factor of $m_ec^2$, the diffusion 
because the electron is moving so slowly (depending on $k_BT/m_ec^2$) that the 
Doppler shift is almost negligible, and the drift is divided by the reduced 
Compton frequency $\omega_c = m_e c^2/\hbar \approx \SI{7.76e20}{\hertz}$ so 
that there is barely any recoil.\\

The derivation of the Kompaneets equation has been called ``distinctly 
non-trivial'' \cite{longair}, and many authors have presented or repeated their 
best derivation.  However, we could not find any reference which carries a full 
derivation  in the same spirit that Kompaneets originally proposed, using the 
correct relativistic expression of the kinetic (Boltzmann) equation. For 
example, it is worth making precise when in fact the diffusion approximation of 
the Boltzmann equation, the standard way of deriving \eqref{ke}, cannot yield a 
continuity equation. We found that such features are not fully explored in the 
existing literature. We believe that a ``more didactical'' derivation  may also 
be useful to highlight physical limitations and possible extensions.  \\  
Kompaneets in 1956 is not explicit about the form of the kinetic equation used 
as starting point\footnote{In his \textit{original} paper \cite{kompa}, 
Kompaneets mentions in a footnote that his work had already been published as 
Report No. 336 at the Institute of Chemical Physics, Academy of Sciences, USSR. 
We exhaustively searched the literature but we could not find this report. 
Interestingly, Peebles mentions as well that he could not find it either 
\cite{peebles}.}.  He proposes to carry an expansion up to second order in the 
photon transferred energy, using the Thomson scattering cross section
\begin{equation}
	\label{thomson}
	\frac{\id \sigma^\text{Th}}{\id \Omega_\text{rest}} = \frac{3\sigma_T}{16\pi}\left(1 + \cos^2\theta_\text{rest}\right)
\end{equation}
where $\id\Omega_\text{rest}$ is the scattering solid angle measured in the electron rest frame.
In 1963 Dreicer \cite{dreicer} more carefully describes the 
Fokker-Planck approximation to the Boltzmann equation for an electron-photon 
gas.  However, he does not give an expression for the time evolution of the 
photon number. In 1965 Weymann \cite{weymann} uses Dreicer's formalism to 
derive an equation identical to the one by Kompaneets, although not displaying 
the details of the calculation. Somewhat remarkably, although Dreicer's and 
Weymann's papers are published after Kompaneets' paper, they do not cite the 
latter. Nevertheless, those as well as many other
references \cite{rybicki, zhang, katz, liu} follow Kompaneets' recipe, also using the Thomson cross section. We highlight that it is not possible to 
find the Kompaneets equation by that approach while performing the diffusion approximation as proposed by Kompaneets; rather we must use the full
relativistic expression with the Klein-Nishina cross section including the 
M\o ller velocity factor (see Section \ref{css}) in order to correct for the flux of electrons in the kinetic 
equation
\eqref{komp-boltz}. In fact, this factor appears naturally while carefully deriving the relativistic Boltzmann equation; see for example \cite{kremer, weert, kolkata}.\\  
	
What follows is split in two parts. Sections \ref{db}--\ref{km} make the first part,  where general considerations  explain how and why the Planck law is obtained from a Master equation for the bosonic occupation in reciprocal space.  By performing a Kramers-Moyal (or diffusion) expansion on the Master equation of a discrete random walk in reciprocal space with suitably chosen transition rates, we  show  that the Kompaneets equation (and its extensions) can be obtained.  That offers a quite different perspective which is especially simple and general. In particular, we believe that this approach is more adapted to carry nonequilibrium extensions on the Kompaneets equation.\\
The second part starts with Section \ref{css} where we connect with the standard setup for deriving the Kompaneets equation. That centers around the diffusion approximation to the Boltzmann equation with the electron gas as stationary background.  It becomes clear then in Section \ref{twointegrals} what are the main steps and what is the nature of the approximation.\\
Apart from talking with history and reviewing part of the existing literature on the derivation 
of the Kompaneets equation, we are motivated  by exploring possible 
extensions.  Our motivation for a better understanding of the Kompaneets physics is based on \cite{arca} where the fascinating 
hypothesis was formulated of a nonequilibrium effect modifying the Planck 
spectrum in the cosmic background radiation.  We do not quite continue that 
search in the present paper, but in Section \ref{gk} we give generalizations and some possible nonequilibrium 
extensions of the Kompaneets equation. 
	  
	  There exist many derivations of the Kompaneets equation, witnessing of its fundamental interest in problems of light-matter interaction.  Revisiting some of them is useful we believe for clarifying subtleties (and even inconsistencies) that are rarely mentioned (and that a careful reader would stumble upon) even in textbook references.  Together with the first part of this paper, a more comprehensive understanding of this famous equation, which is much less known in the statistical mechanics community, pays respect to all these efforts while it may possibly open new avenues to introduce nonequilibrum features in the physics of the Early Universe.
		
\section{Detailed balance with stimulated emission}\label{db}
Pauli was probably the first to dynamically characterize thermal equilibrium 
for photons in an electron bath, identifying in 1923 the condition of detailed 
balance in the Master equation describing scattering and recovering the Planck 
law for its stationary distribution \cite{pauli}.\\
 Transition rates for a jump process arising within a quantum many-particle 
system are derived from one-particle Green's functions \cite{kadanoff}. Here we 
take a photonic set up with a symmetrized Fock space taking the tensor product 
over three-dimensional harmonic oscillators with wave vector ${\mathbf k}$  corresponding 
to frequency $\omega$.  An elementary transition is the annihilation of a 
photon with wave vector ${\mathbf k}$ while creating a photon with wave vector ${\mathbf k'}$.  
When the photons are in weak contact with a thermal bath at inverse temperature $\beta$, each transition creates a flux in reciprocal space with an 
expectation given by
\begin{equation}\label{ww}
j({\mathbf k}\rightarrow {\mathbf k'})= \alpha({\mathbf k},{\mathbf k'})\, 
e^{\beta(\hbar\omega-\hbar\omega')/2}\,\big|\langle \text{f}| 
a^\dagger_{\mathbf k'}a_{\mathbf k}|\text{i}\rangle\big|^2
\end{equation}
where $\alpha({\mathbf k},{\mathbf k'}) = \alpha({\mathbf k'},{\mathbf k})$ is 
symmetric and left unspecified for the moment  as a parameter of dynamical 
activity.   We have taken that
to lowest order  the matrix element will contain a single term annihilating and creating a photon $a^\dagger_{{\mathbf k'}}a_{\mathbf k}$.  Writing $n({\mathbf k})$ for the occupation/level at wave vector ${\mathbf k}$, we put
\[
|\text{i}\rangle = | \ldots n({\mathbf k})\ldots n({\mathbf k'})\ldots\rangle
\]
for the initial state.  The only non-zero matrix element will between that initial $|i\rangle$ and
the final state
\[
|\text{f} \rangle = | \ldots n({\mathbf k})-1\ldots n({\mathbf k'})+1\ldots\rangle
\]
Since on each of the Hilbert spaces $a|n\rangle = \sqrt{n}|n-1\rangle$ and  $a^\dagger|n\rangle = \sqrt{n+1}|n+1\rangle$, we find that
\[
\left|\langle \text{f}\,| a^\dagger_{{\mathbf k'}}a_{\mathbf k}|\text{i}\rangle\right|^2 = (1+n({\mathbf k'}))\,n({\mathbf k})\]
Therefore \eqref{ww} becomes
\begin{eqnarray}\label{www}
j({\mathbf k}\rightarrow {\mathbf k'}) &=& \alpha({\mathbf k},{\mathbf 
k'})\,w({\mathbf k},{\mathbf k'}) \,n({\mathbf k})\\
w({\mathbf k},{\mathbf k'}) &=&  e^{\beta(\hbar\omega-\hbar\omega')/2}\,(1+n({\mathbf k'}))\nonumber
\end{eqnarray}
In classic texts on Markov processes, \eqref{www} makes the sink term into ${\mathbf k'}$ from ${\mathbf k}$ and one would write a Master equation for the probability of occupying the various wave vectors.  The source term is 
$j({\mathbf k'}\rightarrow {\mathbf k})$.  Ignoring however correlations between the occupations at different wave vectors, we can write  the Master equation directly for the (now expected) occupation numbers
\begin{equation}\label{meq}
\frac{\partial}{\partial t} n_t({\mathbf k}) = \sum_{{\mathbf k'}} 
\alpha({\mathbf k},{\mathbf k'})\, \big[ 
e^{\beta(\hbar\omega-\hbar\omega')/2}\,(1+n_t({\mathbf k'}))n_t({\mathbf k}) - 
e^{\beta(\hbar\omega'-\hbar\omega)/2}\,(1+n_t({\mathbf k}))n_t({\mathbf 
k'})\big]
\end{equation}
As such, the evolution equation \eqref{meq} does not need to be photon number-preserving, i.e., it does not follow for free that
\begin{equation}\label{pa}
\sum_{\mathbf k}\frac{\partial}{\partial t} n_t({\mathbf k}) = \sum_{{\mathbf 
k},{\mathbf k'}} \alpha({\mathbf k},{\mathbf k'})[w({\mathbf k'},{\mathbf k}) 
n_t({\mathbf k'}) - w({\mathbf k},{\mathbf k'}) n_t({\mathbf k})] = 0
\end{equation}
unless $\alpha({\mathbf k},{\mathbf k'})=\alpha({\mathbf k'},{\mathbf k})$ is 
indeed symmetric. Only then, \eqref{meq} is a continuity equation.\\
Secondly, detailed balance requires that $j({\mathbf k}\rightarrow {\mathbf 
k'}) - j({\mathbf k'}\rightarrow {\mathbf k})=0$ for all ${\mathbf k},{\mathbf 
k'}$, or (for symmetric $\alpha({\mathbf k},{\mathbf k'})$ always)
\begin{eqnarray}
w({\mathbf k},{\mathbf k'})\,n({\mathbf k}) &=&  w({\mathbf k'},{\mathbf k})\,n({\mathbf k'})\;\;\text{ or }\nonumber\\
 e^{\beta(\hbar\omega-\hbar\omega')/2}\,(1+n({\mathbf k'}))\,n({\mathbf k}) &=&  e^{\beta(\hbar\omega'-\hbar\omega)/2}\,(1+n({\mathbf k}))\,n({\mathbf k'})\nonumber\\
\implies e^{\beta\hbar\omega}\,\frac{n({\mathbf k})}{1+n({\mathbf k})}&=& \text{ constant\,} = e^{\beta\mu}  \end{eqnarray}
which implies that
\[
n({\mathbf k}) = \frac 1{e^{\beta\hbar\omega} -1}
\]
by assuming that the chemical potential equals zero, $\mu=0$.  Note that we have only used \eqref{www} for an environment in thermal equilibrium where the temperature may refer to an electron gas or anything else.  The ``anything else'' would solely show in the prefactor $\alpha({\mathbf k},{\mathbf k'})$ for the kinetics \eqref{meq}.
In that sense the evolution given by \eqref{meq} represents a general 
Kompaneets equation, before any diffusion approximation.

	\section{Kramers-Moyal expansion}\label{km}
	The previous section considers jumps in the space of wave vectors ${\mathbf k}$.  We associate in the present section an energy $U({\mathbf k})$ to the system and we expand the Master equation \eqref{meq} for small energy changes ${\mathbf k}\rightarrow {\mathbf k'}$.  That is more generally known as a Kramers-Moyal or diffusion approximation.\\
Let us start in one dimension, where we consider a lattice mesh $\delta>0$ for $x\in \delta \mathbb{Z}$.  The $x = k_1$ stands for the first component of the (rescaled) wave vector.  We imagine a walker hopping on that lattice of wave vectors, to nearest neighbor sites with transition rates
\begin{eqnarray}
\label{tr}
w(x,x\pm\delta)&&=\\ (1+n(x\pm\delta))B(x \pm 
\frac{\delta}{2})\,&&\exp\left\{-\frac{\beta}{2}\left(U(x\pm\delta)-U(x)\right)\right\}\exp\left\{\frac{\pm\beta\delta}{2}f\left(x\pm\frac{\delta}{2}\right)\right\} \nonumber
\end{eqnarray}
Here, $n$ is the instantaneous number of walkers; its presence in the rates 
represents the stimulated emission. The function $B>0$ is an inhomogeneous 
activity rate and  $\beta$ is the inverse temperature of a medium enabling the 
hopping.  There is also a driving force\footnote{Note we are in reciprocal 
space here so that $\delta$ is an inverse length and $f$ is measured in 
multiples of $\hbar\,c$.} $f$ and a potential $U$ which are added following the 
condition of local detailed balance at fixed environment inverse temperature 
$\beta = 1/k_B T$  \cite{ldb}. We do not for the moment dwell on the physical 
meaning of the driving $f$ and we do not restrict us to photons but to bosonic 
systems more generally; see also Section \ref{gk}.
  The rate \eqref{www} is a special case of \eqref{tr}, where  $f\equiv0$ and the photon energy $U=\hbar\omega$. Abusing notation, we also incorporate the symmetric activity $\alpha({\mathbf k},{\mathbf k'})$ in the rates as the prefactor $B$.\\

For fixed $\delta$ the Master equation as in \eqref{meq} becomes
\begin{eqnarray}
\frac{\partial n_t}{\partial t}(x) &+&  j_t(x,x+\delta) - j_t(x-\delta,x) =0 \;\;\text{ for}\\
j_t(x,x+\delta) &=& n_t(x) w(x,x+\delta) -n_t(x+\delta)w(x+\delta,x) 
\nonumber\\
j_t(x-\delta,x) &=& n_t(x-\delta)w(x-\delta,x) -n_t(x) w(x,x-\delta) \nonumber 
\end{eqnarray}
We expand this last equation to second order in $\delta$; see Appendix \ref{exp}.
The result is
\begin{align}\label{mm}
\frac{\partial}{\partial t}n_t = \delta^2\bigg{\{} &\left(\beta B g' + \beta B'g\right)(1+n)n +\left(\beta B g+B'\right)n' + 2\beta B g n n' + Bn''
\bigg{\}} 
\end{align}
with $g(x) = U'(x)-f(x)$.
That can be written more explicitly as a continuity equation,
\begin{equation}
\frac{\partial n_t}{\partial t}(x) = \delta^2\frac{\partial}{\partial x}\bigg{\{} B(x)\,\bigg(\frac{ \partial n_t}{\partial x}(x)+ \beta g(x)\big(1 + n_t(x)\big)n_t(x) \bigg)
\bigg{\}} 
\end{equation}
in which we recognize the structural elements of the Kompaneets equation \eqref{ke}.\\

We can indeed redo that in three dimensions, on $\delta \mathbb{Z}\times\delta \mathbb{Z}\times\delta \mathbb{Z}$. Taking the same rates in all directions as before with 3-dimensional ``force'' $f$, the diffusion approximation now reads
\begin{align}
\label{serv}
\partial_tn_t = \delta^2\;{\pmb \nabla} \cdot {\mathbf j}
\end{align}
with in Cartesian coordinates $(x_\ell, \ell=1,2,3)$ for ${\mathbf j}=\sum_\ell j_\ell \mathbf{\hat{x}_\ell}$,
\begin{equation}\label{genk}
j_\ell=B\left(\frac{\partial n}{\partial x_\ell}+ \beta g_\ell\,(1 + n)n\right)
\end{equation}
for $g_\ell = \frac{\partial U}{\partial x_\ell} - f_\ell$.\\
Moving finally to the setup for the Kompaneets equation we enter frequency space by assuming that $n_t=n(\omega,t), g=g(\omega), D=D(\omega)$ with frequency $\omega = c \,\sqrt{\sum_\ell x_\ell^2}$ for speed of light $c$.  It means to rewrite \eqref{serv}
in spherical coordinates, with $\omega$ as radial variable: 
\begin{equation}\label{genke}
\omega^2\frac{\partial n}{\partial t}(\omega,t) = c\,\delta^2\, \frac{\partial }{\partial\omega}\left\{\omega^2 B(\omega)\left(c\,\frac{\partial n}{\partial \omega}(\omega,t)+ \beta \,g(\omega)\big(1 + n(\omega,t)\big)n(\omega,t)\right)\right\}
\end{equation}
  That is an extended Kompaneets equation, to be compared with \eqref{ke}, where the energy change and the driving combine into $g(\omega) = c\frac{\partial U}{\partial \omega}(\omega) - f(\omega)$.\\
Making the choices
\begin{align}\label{cho}
c^2\delta^2\,B(\omega)=\frac{k_B T}{m_ec^2 } n_e\sigma_T\,c\,\;\omega^2,\qquad g(\omega) = \hbar\,c
\end{align}
the above equation \eqref{genke} becomes exactly the one of Kompaneets 
\eqref{ke}. Note that $n_e\sigma_Tc= \tau^{-1}$ is the average collision rate, 
as before.  That shows that the full structure of the Kompaneets equation is 	
obtained as the diffusion approximation to a Master equation with stimulated 
emission, and this holds whenever the limiting activity and drift obey 
\eqref{cho}.  Justifications for the choices \eqref{cho} come from the physical 
nature of the process considered in the Kompaneets equation.  The photon energy 
is there $U(\omega) =\hbar \omega$ and there is no driving $f\equiv 0$, making 
indeed $g=c\frac{\partial U}{\partial \omega} = \hbar c$. For understanding the 
first equality in \eqref{cho}, we note that $c^2\delta^2\,B(\omega)$ appears as 
the diffusion constant $D(\omega)$ in \eqref{genke}. The shift in frequency for 
a photon undergoing Compton scattering determines that diffusion constant as 
the conditional average squared shift
\begin{equation}\label{di}
D(\omega) = \left\langle \frac{(\omega' - \omega)^2}{2\tau} \,\bigg|\,\omega\,\right\rangle 
\end{equation}
 The shift follows from the well-known Compton scattering formula \cite{longair} 
\begin{equation}\label{shift}
\omega' - \omega = \frac{c\mathbf{p}\cdot(\mathbf{\hat{n}'} - \mathbf{\hat{n}}) - \hbar\omega (1- 
\mathbf{\hat{n}}\cdot\mathbf{\hat{n}'})}{\gamma m_ec^2 \left[1 - \mathbf{p}\cdot\mathbf{\hat{n}'}/\gamma m_ec + 
(\hbar\omega/\gamma m_e c^2) (1 - \mathbf{\hat{n}} \cdot \mathbf{\hat{n}'})\right]}\, \omega
\end{equation}
where the electron momentum is $\mathbf{p}$, $\gamma$ is the relativistic factor and $\mathbf{\hat{n}'} - \mathbf{\hat{n}}$ is the scattering vector.
In the low-temperature regime where Compton scattering is relevant, under the 
assumption that the electrons and the photons are of comparable energy much 
less than $m_ec^2$, most of the momentum is carried by the electrons, meaning 
$|\mathbf{p}| \gg \omega / c$. Hence we can replace the term between the brackets in the 
denominator by unity, and only retain the first term in the numerator in  
\eqref{shift}
\begin{equation}\label{lin}
\omega' - \omega \approx \frac{\mathbf{p}\cdot(\mathbf{\hat{n}'} - \mathbf{\hat{n}})}{m_e c} \omega
\end{equation}
 We can  assume 
the square of projection of the scattering vector $\mathbf{\hat{n}'}-\mathbf{\hat{n}}$ on the 
momentum vector $\mathbf{p}$ to average out to a constant of magnitude 1, which we 
will hereafter ignore.  Continuing then the calculation for \eqref{di} yields
\begin{equation}
D \propto \left\langle \frac{|\mathbf{p}|^2}{2m_e^2c^2} \frac{\omega^2}{\tau} 
\right\rangle = \frac{1}{m_ec^2} \left\langle \frac{|\mathbf{p}|^2}{2m_e} \right\rangle 
\frac{\omega^2}{\tau} \propto \frac{k_B T}{m_ec^2 } \frac{\omega^2}{\tau}
\end{equation}
where the temperature $T$ gives the average kinetic energy of the electron 
distribution.  We thus recover the first equality in \eqref{cho}.\\

For better understanding and comparison, we add here the heuristics  of the opposite situation where we consider an electron in a photon bath (in contact with other matter at temperature $T$).
Since the electronic density of states goes as 
the square root of the energy $E$, we must have an equation for the number $\cal N(E,t)$ of electrons of the form
\begin{equation}\label{hek}
\frac{\partial \cal N}{\partial t}(E,t)=  \frac{1}{E^{1/2}} \,\frac{\partial 
}{\partial E}E^{1/2}\,j_t(E)
\end{equation}
to be compared with \eqref{cur}.  To determine the current $j_t(E)$ we remember that a crucial property of the Doppler effect, used to compute \eqref{di} and 
to arrive at \eqref{cho}, is that the shift in frequency is linear in the 
frequency itself, as seen in \eqref{lin}. Here however, because the shift 
in the electron's energy equals the same expression, the average squared energy 
shift is proportional to the square of the momentum instead, hence only 
linear in the energy.  That allows us to write down the electronic 
version of the Kompaneets equation immediately
\begin{equation}\label{ek}
E^{1/2} \,\frac{\partial \cal N}{\partial t}(E,t)=  b\,\frac{\partial 
}{\partial E}E^{3/2}\left\{k_B T \frac{\partial \cal N}{\partial E}(E,t) + \cal N(E,t)\right\}
\end{equation}
for some rate $b \propto  c\sigma_T\,\frac{U_\gamma}{m_e c^2}$, where $U_\gamma$ is the energy density of the photon gas. Note finally that we neglect the fermionic nature of the electron, since we presume non-degeneracy of the electron gas (not having thus the fermionic version of stimulated emission).   For verification, a derivation of this 
equation can be found in \cite{electronkompaneets}. For applications to highly dense states of fermionic matter, the Pauli exclusion is significant and departures from \eqref{ek} are expected. In such a regime however, the long-range Coulomb interactions become relevant and an extra term must be added to the Boltzmann equation \eqref{ek} anyway. 

\section{The Boltzmann-Master equation }\label{css}
The Master equation \eqref{meq} only involves the photon occupation and the 
electron bath is integrated out and remains present only via the bath temperature. 
Similarly, the dynamics \eqref{tr} effectively treats the electron bath via 
temperature, mobility and possible driving $f$. To go back one (finer) level of 
description, we must introduce the integration step. Here we start from a 
Master equation jointly for electrons and photons. Formally, it has the general 
structure\footnote{From here we make a distinction between the vectors $\mathbf p$ etc and the four-vectors $p$.},
\begin{alignat}{2}
\label{komp-mas}
\frac{\partial n}{\partial t}(\mathbf p,\mathbf k;t) &= \int \id^3\mathbf{p}\,' 
\,\id^3\mathbf{k}' &\{&n(\mathbf p\,',\mathbf k';t)w(p\,',k'\rightarrow 
p, k)(1+ n(\mathbf k,t))\nonumber\\
&&- &n(\mathbf p,\mathbf k;t)w( p, k\rightarrow  p\,',k')(1+ n(\mathbf 
k',t))\}
\end{alignat}
for the joint occupation $n(\mathbf p,\mathbf k;t)$ at time $t$ of electron momentum $\mathbf 
p$ and photon wave vector $\mathbf k$. The four-momenta participating in the transition are, of course, given by
\[
p^{(')} = \left(\frac{E}{c}^{(')},\mathbf{p}^{(')}\right) \, \ \mathrm{and} \ \ k^{(')}=\left(\frac{\hbar \omega}{c}^{(')}, \mathbf{k}^{(')}\right)
\]
We added the stimulated emission in terms 
of the photon occupation $n(\mathbf k',t)$, but include no such factor for the 
electrons, assuming them to be nondegenerate (dilute). The transition rates 
should be obtained from the microscopic process of interaction, here Compton 
scattering, but we write generally
\begin{equation}\label{rates}
w(p, k\rightarrow  p\,', k')\,\id^3\mathbf{p}'\, 
\id^3\mathbf{k}' = \frac{\id \sigma}{\id \Omega}(\mathbf p,\mathbf 
k)\,\id \Omega\,c\,\left(1- \frac{\mathbf v}{c} \cdot \mathbf{\hat{n}}\right)
\end{equation}
for a differential cross section $\frac{\id \sigma}{\id \Omega}$, where the 
equality holds in the sense of distributions.  The factor $c(1-\mathbf{v}/c \cdot 
\mathbf{\hat{n}})$ is commonly called the \textit{M\o ller velocity} and it is necessary for a consistent microscopic description \cite{kremer,mirco,kolkata}. Neglecting that factor 
is a common inaccuracy in the literature, e.g. \cite{electronkompaneets, chen, 
tong}. Viewed nonrelativistically it corresponds to the relative velocity 
between the electrons and the photons, but even with a proper relativistic 
treatment, where the relative velocity is of course $c$ in every frame, this 
exact factor will show in the colliding flux density \cite{terrall}.\\

We want \eqref{komp-mas} to correspond to a Boltzmann equation where the total 
number of electrons and photons is conserved. It suffices here to invoke 
dynamical reversibility in the form
\begin{equation}\label{dyn}
w(p,k\rightarrow p\,', k') = w( p\,', k'\rightarrow 
p, k)
\end{equation}
which means that the rate function\footnote{To be consistent with literature, 
we note here that $w( p, k\rightarrow  p\,', k')$ defined as such is called the 
\textit{non-covariant transition rate}. The \textit{covariant transition rate} 
$W( p, k\rightarrow  p\,', k')$ is defined such that $w( p, k\rightarrow  p\,', 
k') = \frac{W( p, k\rightarrow  p\,', k')}{EE'\hbar \omega \hbar\omega'}$} is 
symmetric. Moreover, if we assume that the electron gas evolves on much shorter 
time-scales than the photons, we can suppose to two distributions are 
uncorrelated, allowing us to write
\[
n(\mathbf p,\mathbf k;t) = \cal N(\mathbf p)\,n(\mathbf k, t),\]
where $\cal N(\mathbf p)$ is the momentum distribution of the electrons, which 
need not necessarily be in thermal equilibrium. We also assume isotropy so that 
the above leads to the writing of the Boltzmann-type equation:
\begin{equation}
\label{komp-boltz}
\frac{\partial n}{\partial t}(\omega) =  \int \id^3\mathbf{p} \,\id w\, 
\left[n(\omega') {\cal N}(\mathbf{p'}) (1+n(\omega)) - n(\omega) {\cal N}(\mathbf{p}) 
(1+n(\omega')) \right] \end{equation}
where $\id w= \id \sigma\, c(1- \mathbf v / c\cdot\mathbf{\hat{n}}), \id \sigma = \frac{\id 
\sigma}{\id \Omega}\,\id\Omega$ stands for the (unspecified for the moment) 
scattering cross section multiplied with the incoming flux of electrons.
The scattering is subject to conservation of energy-momentum corresponding to the collision scheme
\[
\left(p \,,\, k\right)\, \rightleftharpoons\, \left(p' \,,\, k'\right)
\]
which means a transition of wave vectors and corresponding frequencies
\begin{equation}
\mathbf{k}  \rightleftharpoons \mathbf{k}^{'},\quad  \omega = c|\mathbf k| \rightleftharpoons  \omega' = c|\mathbf k'| 
\end{equation}
for the photon. 
The first term on the right-hand side of \eqref{komp-boltz} is the source term where photons with frequency $\omega$ are created from collisions with electrons with momentum $\mathbf p'$, and the second term is the sink term where photons with frequency $\omega$ collide with electrons of momentum $\mathbf p$.  In that case, the equation \eqref{komp-boltz} is rewritable indeed as a Master (rate) equation and is really the more precise version of Eq.~1 in \cite{kompa}.\\

We can 
integrate out the electron bath to find the rates involving only the photons
\begin{equation}
\label{in2}
w(k'\to  k)= \int \id^3\mathbf{p} \, \id^3\mathbf{p}\,' \, w( p', k'\rightarrow 
p,k)\mathcal{N}(\mathbf p').
\end{equation}
Invoking the reversibility \eqref{dyn} of the rates, $w(p',k'\to p, k) = w(p, k 
\to p',k')$, we continue by writing down the inverse rates
\begin{equation}
\label{in3}
w(k\to k')= \int \id^3\mathbf{p} \, \id^3\mathbf{p}\,' \, w(p',k'\rightarrow 
p,k)\mathcal{N}(\mathbf p')\frac{\mathcal{N}(\mathbf p)}{\mathcal{N}(\mathbf p')}
\end{equation}
To connect this to the setup in Section \ref{db}, we have to further include 
the assumption that the electrons are in thermal equilibrium with inverse 
temperature $\beta = 1/k_B T$, which implies that 
\begin{equation}\label{conserve}
\frac{\mathcal{N}(\mathbf p)}{\mathcal{N}(\mathbf p')} = e^{-\beta(E - E')} = 
e^{-\beta(\hbar\omega' -\hbar \omega)} ,\end{equation} where the last equality 
follows from conservation of energy in the collisions. Inserting 
\eqref{conserve} into \eqref{in3} yields the detailed balance relation for the photon transition rates \eqref{in2},
\[
\frac{w(k'\to k)}{w(k\to k')} = e^{-\beta(\hbar \omega - \hbar \omega')}.
\]

\section{Diffusion approximation to the Boltzmann Equation}\label{twointegrals}

The previous section shows how the kinetic equation which correctly introduces the Compton scattering between the electrons (making the steady thermal bath) and the photons gives rise to a Master equation as treated in Section \ref{db}. From there, as shown in Section \ref{km}, the Kompaneets equation can be obtained.  In the present section, we take the two steps together and directly apply the diffusion approximation to the Boltzmann equation.  It is much closer to the published approaches and it gives us the opportunity to clarify a number of points which traditionally are left to the reader. In particular, our framework is more direct than \cite{kompa,rybicki,dreicer,weymann} because we do not \emph{use} (in the derivation itself) that the Planck distribution is the stationary solution of \eqref{komp-boltz}. Under certain conditions on the possible cross sections, Escobedo and Mischler \cite{escobedo1} rigorously prove the diffusion approximation to the Boltzmann equation in the context of an electron-photon gas. However, they do not explicitly link these conditions to the Thomson/Klein-Nishina cross section nor do they address the same problems we treat here. \\

We begin with \eqref{komp-boltz} where $\id w = \id\Omega \,c\left(1 -\frac{ \mathbf{v}}{c}\cdot\mathbf{\hat{n}}\right)  \frac{\id\sigma}{\id \Omega}(\mathbf{p},\mathbf{\hat{n}}, \Omega)$ and
\begin{align}
\label{klein-nishina}
\frac{\id\sigma}{\id \Omega}(\mathbf{p},\mathbf{\hat{n}}, \Omega) = &\frac{3 \sigma_T}{16 \pi}\frac{1}{\gamma^2 \left(1- \mathbf{p}\cdot \mathbf{\hat{n}'}/\gamma m_ec + \frac{\hbar\omega}{\gamma m_ec^2} (1-\mathbf{\hat{n}}\cdot\mathbf{\hat{n'}}) \right)^2}\times \nonumber \\ \nonumber
  &\Bigg\{ 1 +\frac{(\frac{\hbar\omega}{\gamma m_ec^2} (1-\mathbf{\hat{n}}\cdot\mathbf{\hat{n'}}))^2}{(1- \mathbf{p}\cdot \mathbf{\hat{n}'}/\gamma m_ec)(1- \mathbf{p}\cdot \mathbf{\hat{n}'}/\gamma m_ec + \frac{\hbar\omega}{\gamma m_ec^2} (1-\mathbf{\hat{n}}\cdot\mathbf{\hat{n'}}))}\ + \\ 
  & \ \ \ \  +\left(1 - \frac{(1-\mathbf{\hat{n}}\cdot\mathbf{\hat{n'}})}{\gamma^2( 1 - \mathbf{p}\cdot \mathbf{\hat{n}}/\gamma m_ec)(1 - \mathbf{p}\cdot\mathbf{\hat{n}'}/\gamma m_ec))}\right)^2\Bigg\}
\end{align}
as the (correct) Klein-Nishina expression for the cross section. This expression appears in \cite{barbosa}, but a derivation is not shown there. Jauch and Rohrlich \cite{jauch} give a detailed derivation, but only express the differential cross section in terms of the scattering matrix. As we could not find any reference which contains the exact frame-independent expression for the Klein-Nishina cross section given above we will devote Appendix \ref{knder} to this discussion. However, we also note here that once established the correct (and not so enlightening) expression above, we only need the first few orders of its expansion. As before, we keep in mind the collision scheme $$\mathbf{p} + \frac{\hbar\omega}{c}\mathbf{\hat{n}} \rightleftharpoons \mathbf{p'} + \frac{\hbar \omega'}{c}\mathbf{\hat{n}'}$$ where we now express the wave vector as $\mathbf k = \frac{\hbar \omega}{c}\mathbf{\hat{n}}$, defining the scattering angle $\cos\theta = \mathbf{\hat{n}}\cdot \mathbf{\hat{n}'}$.\\

We follow the usual steps, assuming that: (i) the electrons are in thermal equilibrium at temperature $T$; (ii) the photons are soft, meaning that their energy is very small compared to the rest energy of the electron ($\hbar\omega \ll m_ec^2$), but of same order of the electron bath energy ($\hbar\omega \sim k_BT$) ; (iii) electrons are nonrelativistic ($|\mathbf{p}|\ll m_ec$ or $k_BT\ll m_ec^2$). By combining (ii) and (iii) we conclude also that energy is transferred in small amounts only, permitting the continuum (or diffusion) approximation\footnote{In fact, the transfer of energy depends both on the incoming $\omega$ and $\mathbf{p}$ as dictated by the Compton formula and, to the lowest order, it is proportional to the product of both \eqref{lin}.}.\\
Points (ii) and (iii) suggest an expansion in terms of the energy shift, as proposed originally by Kompaneets \cite{kompa}. For this purpose we define
\[
\Delta = \frac{\hbar(\omega' - \omega)}{k_B T}
\]
By looking at the Compton shift \eqref{shift}, we have
\begin{equation}
\label{cs}
    \Delta(\omega, \mathbf{p}) = \frac{\hbar \omega}{k_B T}  \frac{c\mathbf{p}\cdot(\mathbf{\hat{n}'} - \mathbf{\hat{n}}) - \hbar\omega (1- 
    	\mathbf{\hat{n}}\cdot\mathbf{\hat{n}'})}{\gamma m_ec^2 \left[1 - \mathbf{p}\cdot\mathbf{\hat{n}'}/\gamma m_ec + 
    	(\hbar\omega/\gamma m_e c^2) (1 - \mathbf{\hat{n}} \cdot \mathbf{\hat{n}'})\right]} 
\end{equation}
We refer the reader to \cite{jauch} for more discussion of expressions \eqref{klein-nishina} and \eqref{cs}.\\

Let us now go back to \eqref{komp-boltz} to make an expansion up to second order in $ \Delta(\omega, \mathbf{p})$, mimicking the Kramers-Moyal expansion of Section \ref{km}. We make the natural change of variables
\begin{align*}
	&\omega \to x= \frac{\hbar \omega }{k_B T}\\
	&\omega'\to x'= \frac{\hbar \omega' }{k_B T}
\end{align*}
 to write the photon numbers up to second order as
\begin{align}
\label{pdist1}
&n(x',t)( 1 + n(x,t)) = n(x,t)(1+n(x,t)) + (1+n(x,t))\frac{\partial n}{\partial x} \ \Delta + (1+n(x,t))\frac{\partial^2 n}{\partial x^2} \ \frac{\Delta^2}{2}\\
\label{pdist2}
&n(x,t)( 1 + n(x',t)) = n(x,t)(1+n(x,t)) +  n(x,t)\frac{\partial n}{\partial x} \ \Delta + n(x,t)\frac{\partial^2 n}{\partial x^2} \ \frac{\Delta^2}{2}
\end{align}
From (i) above, the electron distribution is Maxwellian at temperature $T$,
\begin{equation}
\label{maxwell}
    {\cal N}(\mathbf{p})\id^3\mathbf{p} = {\cal N}_\text{eq}(|\mathbf{p}|)\id^3\mathbf{p} = n_e (2\pi m_e k_B T)^{-3/2}\exp\left(-\frac{p_x^2 + p_y^2 + p_z^2}{2m_ek_BT}\right)\id^3\mathbf{p}
\end{equation}
where $n_e$ is the electron density. Using conservation of energy $$E' = E - \Delta \ k_BT$$ we get
\begin{equation}
\label{edist}
    {\cal N}_\text{eq}(|\mathbf{p}'|) = {\cal N}_\text{eq}(|\mathbf{p}|)\left(1 + \Delta + \frac{\Delta^2}{2}\right)
\end{equation}
By taking \eqref{pdist1}--\eqref{pdist2} and \eqref{edist} in \eqref{komp-boltz}, we obtain
\begin{align}
\label{kwi}
    \frac{\partial n}{\partial t} =\left[\partial_x n + n(1+n)\right]I_1 +\left[\frac{1}{2}\partial_{xx} n + (1+n)\left(\frac{n}{2} + \partial_x n\right)\right]I_2 \
\end{align}
where, with $\ell=1,2$,
\begin{equation}
I_\ell(x) = c\int \id^3\mathbf{p}\, \id\Omega \left(1 -\frac{ \mathbf{v}}{c}\cdot\mathbf{\hat{n}}\right) \frac{\id\sigma}{\id \Omega}(\mathbf{p},\mathbf{\hat{n}}, \Omega) \, {\cal N}_\text{eq}(|\mathbf{p}|) \Delta^\ell
\end{equation}
The computation of the integrals $I_1(x)$ and $I_2(x)$ is presented in 
Appendices \ref{int1} and \ref{int2}, with results
\begin{align}
\label{firi}
 &I_1(x)= \frac{n_e\sigma_T c\, k_BT}{m_ec^2}\;x(4-x)\\
\label{seci}
 &I_2(x)= \frac{n_e\sigma_T c\, k_BT}{m_ec^2}\;2x^2
\end{align}
After standard manipulations, we end up with the Kompaneets equation
\begin{equation}
 \label{kompfinal}
    \frac{\partial n}{\partial t}(x,t) = \frac{n_e\sigma_T c\, k_BT}{m_ec^2}\frac{1}{x^2}\partial_x\left\{x^4(\partial_xn(x,t) +n(x,t)(1+n(x,t)))\right\}
\end{equation}
which is \eqref{ke}.\\

As already mentioned, the common setup in the literature is to consider the Thomson cross section \eqref{thomson} \emph{only}, without the M\o ller velocity factor 
\begin{equation*}
	c\left( 1- \frac{\mathbf{v}}{c}\cdot \mathbf{\hat{n}}\right)
\end{equation*}
to compute $I_2(x)$; the value of $I_1(x)$ is then obtained from assuming that \eqref{kwi} must be a continuity equation, with the current vanishing in equilibrium for the Planck distribution. However, when using the Thomson cross section \textit{only} one finds (see Appendix \ref{intthomson} for the computation of the integrals in this case)

\begin{equation}
\label{firiwrong}
I^\text{Th}_1(x)= \frac{n_e\sigma_T c\, k_BT}{m_ec^2}x(1-x)
\end{equation}
instead of \eqref{firi}. To the best of our knowledge that first Kompaneets 
integral was never computed. As a matter of logic, as we have seen in the 
previous section, it is not possible with solely the Thomson cross section 
(which ultimately leads to \eqref{firiwrong}) to find \eqref{kompfinal}.  The 
M\o ller velocity factor must be included in the definition of the rates, 
otherwise consistency in the microscopic description of the collision term is 
lost. This term naturally appears when carefully deriving \eqref{komp-boltz}, 
see \cite{kremer}, or when going from the covariant description to the 
non-covariant one, \cite{kolkata}. As mentioned, this term is also needed to 
yield conservation of the photon number, but what is surprising perhaps, is 
that none of these effects are seen on $I_2(x)$, whence making it possible to 
employ the traditional indirect argument used traditionally (including in 
Kompaneets' original paper) to fix the value of $I_1(x)$.\\
It is possible in the derviation of the Kompaneets equation to use the Thomson 
cross section together with the M\o ller factor (of course) as it satisfies 
dynamical reversibility as well\footnote{In fact, any cross section which is 
obtained from an unitary scattering matrix yields dynamical reversibility; see 
\cite{kolkata} for a proof.}. Note however that upon writing \eqref{thomson} we 
are fixing the electron rest frame and, thus, all quantities must be expressed 
in this frame of reference (including the distribution functions). The 
diffusion approximation and the integrals must also be performed in this frame. 
We can avoid this complication by expressing the cross section in a 
frame-independent manner, just as \eqref{klein-nishina}. In that way we have 
the freedom to work with the Boltzmann equation \eqref{komp-boltz} in the most 
convenient frame of reference\footnote{\textit{Non-covariance} should not be 
confused with \textit{frame-independence}. \eqref{komp-boltz} is an example of 
a non-covariant equation which is written in a frame-independent way.} (where the electron distribution is isotropic Maxwellian). We could not find in the 
literature any frame-independent version of the Thomson cross section, neither 
a mention of such features which are so important to yield a consistent 
diffusion approximation.\\

Traditionally, equation \eqref{komp-boltz} is called the \textit{non-covariant Boltzmann equation} and while a fully covariant Boltzmann equation does indeed exist, we have chosen to not express it here. However, we invite the reader to check, for example, \cite{kolkata, brown} for the covariant formalism. These two versions are equivalent to each other, provided that we use the rates \eqref{rates} to connect them. On the other hand, once the link is established, we can regard the Kompaneets equation as the nonrelativistic limit of either Boltzmann equation, which here is concretely realized as the second-order expansion in the energy shift, as expressed in the integrals \eqref{firi} and \eqref{seci}. In this more kinematic approach, the interpretation of both integrals above is clear as the average shift and shift-squared of the energy, while in covariant approaches\footnote{For simplicity, we call any derivation of the Kompaneets equation starting from the covariant Boltzmann equation a \textit{covariant approach/derivation}. Strictly speaking, the use of the word \textit{covariant} is not entirely correct because when assuming isotropy to carry the diffusion approximation, one is fixing the reference frame, thus breaking covariance.} the interpretation is not always straightforward.
Derivations of the Kompaneets equation which proceed in a covariant manner include \cite{brown, itoh, cooper}.  The covariant derivation is simpler in the sense that the main ingredient is the microscopic reversibility of the rates, in the spirit of Sections \ref{db}--\ref{km}.  The problems we mention above are taken care of automatically then.  For example, using the scattering matrix gives the M\o ller prefactor and consistency of the different reference frames (since the quantities are always expressed in a Lorentz invariant way). In \cite{cha}, Challinor and Lasenby managed to identify the moments of the shift (as above) in a covariant approach, and that was repeated later in \cite{itoh}. Yet,  in the integrals, the  moments are still multiplied by a Dirac delta, which leaves open the exact analytical computation.

	\section{Extending the Kompaneets equation}\label{gk}
	There are various well-known extensions of the Kompaneets equation compared to what was mainly discussed in the previous sections.  For example, the condition that the photons are soft, \textit{i.e.}, $\hbar \omega \sim k_B T$ may be relaxed provided that we treat the regime $\hbar \omega \gg k_B T$ (hard photons) explicitly in the diffusion approximation to the Boltzmann equation.  An extra contribution is found then in \eqref{ke}, referred to as the ``extended Kompaneets equation to down-Comptonization'' \cite{zhang,liu}. The regime of down-Comptonization first appeared in \cite{ross} in the so-called Ross-McCray equation, where radiative transfer of X-ray photons is treated. The equation derived in \cite{ross}, however, does not yield the Bose-Einstein distribution as stationary solution and should be regarded only as an asymptotic limit of such extended Kompaneets equation.\\
	There are more and different processes in astrophysical plasmas that change the photon occupation number. Beyond Compton scattering, contributions due to (for example) {\it Bremsstrahlung} (involving the ions as well) and radiative (or double) Compton scattering may be considered, \cite{kompa, hu, longair, rybicki}. Due to the radiative nature of such processes it is not true any longer that the photon number is conserved and we cannot describe them by a Master-Boltzmann equation.\\
	 Relativistic 
	extensions to the Kompaneets equation and calculations done in the covariant 
	Boltzmann description include \cite{brown, itoh, itoh2, cooper, kohyama1, kohyama2, kohyama3}. Barbosa \cite{barbosa} 
	also addresses relativistic extensions, but using a more kinematic approach to 
	the Fokker-Planck approximation.\\
	Pitrou \cite{pitrou} and Buet {\it et al} \cite{buet} both deal with anistropic extensions to the Kompaneets equation, where the assumption of isotropy in the distribution functions is no longer required. More general solutions to \eqref{ke} can be found in \cite{escobedo2, mihu}.  More recently, the diffusion approximation as proposed by Kompaneets is applied to a neutrino gas \cite{suwa, wang}.\\
	For applying the Kompaneets equation in the context of cosmology, the contribution of the cosmic expansion (red shift) must be taken into consideration.  In \cite{hu, burigana1, burigana2, bernstein}, that contribution appears as a convective term in the Boltzmann equation \eqref{komp-boltz}.  A turbulent Doppler shift (as an integrated Sachs-Wolfe effect for a random gravitational potential field) would give an additional diffusion in frequency but we have not seen that being worked out.

	Let us now turn to less standard generalizations.
Recall that we have already derived an extended Kompaneets equation in 
\eqref{genke} for bosons with energy $U$ and possibly driven by the ``force'' $f$.  Here we continue with a modification and two extra constraints where the electron bath need not be in equilibrium. 
As already pointed out in \cite{barbosa, brown,brown2, peebles}, Kompaneets 
equation can indeed be recovered for an electron bath which is not in 
equilibrium, provided that the distribution of the electrons is isotropic, 
\textit{i.e.}, that the distribution can be written in terms of the energy.  
Although some references already considered this fact by using different 
approaches, including \cite{barbosa, brown}, we could not find references which 
conclude the same while starting from the non-covariant kinetic equation 
\eqref{komp-boltz}. The details of such a derivation are given in Appendix 
\ref{intnoneq}.  By assuming: (i) isotropy of the distribution of the 
electrons, \begin{equation}
 {\cal N}(\mathbf{p},t)\id^3\mathbf{p} = {\cal N}(|\mathbf{p}|,t)\id^3\mathbf{p}
\end{equation}
and (ii) that ${\cal N}$ decays faster than $|\mathbf{p}|^3$, \textit{i.e.},
 \begin{equation}\label{limit}
 \lim_{|\mathbf{p}|\to \infty}|\mathbf{p}|^3{\cal N}(|\mathbf{p}|,t)=0
\end{equation} 
we find Kompaneets equation \eqref{kompfinal} with an effective temperature $$T_\text{eff}= \frac{\langle |\mathbf{p}|^2 \rangle}{3 k_B m_e}$$ where $$n_e\,\langle |\mathbf{p}|^2\rangle =\int \id^3\mathbf{p}\, |\mathbf{p}|^2\, {\cal N}(|\mathbf{p}|)$$
Note that it is not so strange to recover the equilibrium (relaxation to) Planck distribution in the diffusion approximation.  It only means that nonequilibrium features hide in higher order approximations.  As possible scenarios for generating nonequilibrium effects, we may have in mind gravitational sources (still largely unspecified) or the possible presence of (turbulent) magnetic fields driving the electrons.

	\section{Conclusions and outlook}\label{con}
Relaxation to quantum equilibrium distributions is a topic of much current interest. In that spirit, the Kompaneets equation  for the Comptonization of photons in an electron bath is one of the few examples where a diffusion approximation to the Boltzmann equation can be explicitly done, in such way that we found useful to revisit its derivation.\\

First, we connected the derivation with statistical mechanical setups, known for example as Kramers-Moyal expansion and clarified a number of points related to the importance of using the physically correct scattering cross sections together with an \textit{ab initio} correct Boltzmann equation \eqref{komp-boltz}. We explored and stressed many features which are absent or commonly neglected in literature, such as the appearance of the M\o ller velocity in the Boltzmann collision term.

Secondly, the re-derivations in the present paper may serve as useful point of departure to understand modifications and nonequilibrium corrections to the Planck law.  Indeed, one of the goals of the present paper has been to derive the Kompaneets equation without resorting to the Kompaneets' approach based on the current vanishing at equilibrium. We found that diffusion in frequency space, with drift determined by exchanges in energy and possibly nonconservative processes combined with stimulated emission, leads to an extension of the Kompaneets equation. Such extensions will be used in future work to motivate specific low-frequency modifications to the Planck law, following the scenario and motivation in \cite{arca}.

%

\appendix
	
    \section{Expansion of the transition rates}\label{exp}
    The transition rates \eqref{tr}  can be expanded from
    \begin{align*}
    &w(x,x\pm\delta)= \left(1+ n \pm\delta n' +\frac{\delta^2}{2}n''\right)\left(B \pm\frac{\delta}{2}B' + \frac{\delta^2}{8}B''\right)\exp\left\{-\frac{\beta}{2}\left(\pm\delta (U'-f) +\frac{\delta^2}{2}(U''-f')\right)\right\}\\
    &w(x\pm\delta, x)= \left(1+n\right)\left(B \pm\frac{\delta}{2}B' + \frac{\delta^2}{8}B''\right)\exp\left\{\frac{\beta}{2}\left(\pm\delta (U'-f) +\frac{\delta^2}{2}(U''-f')\right)\right\}
    \end{align*}
An expansion up to second order in $\delta$ yields
\begin{align}
&w(x,x\pm\delta)= A(x) \pm \frac{\delta}{2} C(x) + \frac{\delta^2}{2} E(x)\\
&w(x\pm \delta, x)=A(x) \pm \frac{\delta}{2} F(x) + \frac{\delta^2}{2} G(x)
\end{align}
with short-hands
\begin{align*}
&A(x) = B(1+n)\\
&C(x) = 2Bn' - (\beta B g - B')(1+n)\\
&E(x) = \left(\frac{1}{4}(\beta^2B g^2+B'')-\frac{1}{2}(\beta g B' + \beta g' B)\right)(1+n) - (\beta B g - B')n' + Bn'' \\
&F(x) = (\beta B g + B')(1+n)\\
&G(x) = \left(\frac{1}{4}(\beta^2Bg^2 + B'')+\frac{1}{2}(\beta g B' +\beta g'B)\right)(1+n)\\
&g(x)= U'(x)-f(x)
\end{align*}
That gives to leading order in the Master equation
	\begin{equation}
	\partial_tn = \delta^2\left\{(G(x)-E(x))n(x) + F(x) \partial_xn(x) + A (x) \partial_{xx} n(x)\right\}
	\end{equation}
	Substituting the short-hands we get \eqref{mm}.
	
	\section{Klein-Nishina differential cross section}\label{knder}
   
   As before, let us suppose a collision process given by
   \begin{equation*}
   	(p,k) \leftrightharpoons (p',k')
   \end{equation*}
	with four-momenta
	\[
	p^{(')} = \left(\frac{E}{c}^{(')},\mathbf{p}^{(')}\right) \, \ \mathrm{and} \ \ k^{(')}=\left(\frac{\hbar \omega}{c}^{(')}, \mathbf{k}^{(')}\right)
	\]
	we fix our metric tensor to $\mathrm{diag}(- + + +)$ and use the same notation as Jauch and Rohrlich \cite{jauch}. For convenience, we will use natural units ($\hbar=c=1$), which will be restored later in \eqref{klein-nishina}. The total Klein-Nishina cross section for Compton scattering is obtained directly from the trace of the scattering matrix (the scattering amplitude), which is given by
	\begin{equation}
		\sigma =\frac{1}{-p'\cdot k'}\frac{1}{E'\omega'} \frac{3\sigma_T}{16 \pi} \int \id^3 \mathbf{p}' \id\omega' \id\Omega \omega'^2 \delta^{(4)}(p'+k'- p -k) \overline{X}
	\end{equation}
   where $\overline{X}$ is the scattering amplitude. For a proof of this relation we refer the reader to \cite{jauch, cross, cross2}.
   By differentiating with respect to the solid angle and using the Fundamental Theorem of Calculus we obtain the differential cross section
   \begin{equation*}
   	\frac{\id \sigma}{\id \Omega} = \frac{1}{-p'\cdot k'}\frac{1}{E'\omega'}\frac{3\sigma_T}{16\pi} \int \id^3 \mathbf{p}' \id\omega'\omega'^2 \delta^{(4)}(p'+k'- p -k) \overline{X}
   \end{equation*}
this is the frame-independent representation of the differential cross section, where the frame-independent representation of the scattering angle is given by $\cos\theta=\mathbf{\hat{n}}\cdot\mathbf{\hat{n'}}$. By evaluating the integrals we get the following expression for the Klein-Nishina differential cross section
   \begin{equation}\label{kn1}
   	\frac{\id \sigma}{\id \Omega} = \frac{3\sigma_T}{16\pi}\left(\frac{\omega '}{\omega}\right)^2 \frac{1}{\gamma^2(1-\pmb{\beta}\cdot \mathbf{\hat{n}})^2}\overline{X}
   \end{equation}
where $\pmb{\beta}=\mathbf{v}/c$ (or $\mathbf{v}$ in natural units). The outgoing photon frequency $\omega'$ is dependent from the ingoing frequency by the Compton shift \eqref{shift}, which can be expressed in the following way
   \begin{equation}\label{shift2}
   	\frac{\omega '}{\omega } = \frac{1-\pmb{\beta}\cdot \mathbf{\hat{n}}}{1-\pmb{\beta}\cdot\mathbf{\hat{n}'} + \frac{\omega}{E}(1-\mathbf{\hat{n}}\cdot\mathbf{\hat{n}'})}
   \end{equation}
The scattering amplitude is given by
   \begin{equation}\label{samp}
   	\overline{X} = \frac{p\cdot k }{p\cdot k'} + \frac{p\cdot k '}{p\cdot k} + 2\left(\frac{m^2}{p\cdot k'} - \frac{m^2}{p\cdot k}\right) + \left(\frac{m^2}{p\cdot k'} - \frac{m^2}{p\cdot k}\right)^2   
   \end{equation}
   Substitute \eqref{shift2} in \eqref{kn1} to yield
   \begin{equation}\label{kn2}
   	\frac{\id \sigma}{\id \Omega} = \frac{3\sigma_T}{16\pi}\left[\frac{1}{\gamma(1-\pmb{\beta}\cdot \mathbf{\hat{n}'} + \frac{\omega}{E}(1-\mathbf{\hat{n}}\cdot\mathbf{\hat{n'}}))}\right]^2\overline{X}
   \end{equation}
   In \eqref{samp}, using $p\cdot k^{(')} = -E\omega^{(')}(1-\pmb{\beta}\cdot\mathbf{\hat{n}^{(')}})$ we have
   \begin{align*}
   	&\frac{p\cdot k }{p\cdot k'} + \frac{p\cdot k '}{p\cdot k} = 1 + \frac{\omega(1-\mathbf{\hat{n}}\cdot\mathbf{\hat{n}'})}{E(1-\pmb{\beta}\cdot\mathbf{\hat{n'}})} + \frac{1-\pmb{\beta}\cdot\mathbf{\hat{n}'}}{1-\pmb{\beta}\cdot\mathbf{\hat{n}} + \frac{\omega}{E}(1-\mathbf{\hat{n}}\cdot\mathbf{\hat{n}'})}\\
   	&\frac{m^2}{p\cdot k'} - \frac{m^2}{p\cdot k} = - \frac{1-\mathbf{\hat{n}}\cdot\mathbf{\hat{n}'}}{\gamma^2(1-\pmb{\beta}\cdot\mathbf{\hat{n}})(1-\pmb{\beta}\cdot\mathbf{\hat{n}'})}
   \end{align*}
   Substituting both expressions into \eqref{samp} yields
   \begin{equation}\label{samp2}
   	\overline{X} = 1 + \left[ 1-\frac{(1-\mathbf{\hat{n}}\cdot\mathbf{\hat{n}'})}{\gamma^2(1-\pmb{\beta}\cdot\mathbf{\hat{n}})(1-\pmb{\beta}\cdot\mathbf{\hat{n}'})} \right]^2 + \frac{\omega^2(1-\mathbf{\hat{n}}\cdot\mathbf{\hat{n}'})^2}{E^2(1-\pmb{\beta}\cdot\mathbf{\hat{n}'})(1-\pmb{\beta}\cdot\mathbf{\hat{n}'} + \frac{\omega}{E}(1-\mathbf{\hat{n}}\cdot\mathbf{\hat{n}'}))} 
   \end{equation}

   Putting back \eqref{samp2} into \eqref{kn2} we use 
   $$\pmb{\beta}=\frac{\mathbf{p}}{E}$$
    to yield the frame-independent representation of the Klein-Nishina cross section \eqref{klein-nishina}.

	\section{Computation of integrals}\label{compint}
	\subsection{Integral $I_1$}\label{int1}
	We compute 
	\begin{eqnarray*}
	I_1(x) = c\int \id^3\mathbf{p} \ \id\Omega\left(1 -\frac{ \mathbf{v}}{c}\cdot\mathbf{\hat{n}}\right)\frac{\id \sigma}{\id \Omega}(\mathbf{p},\mathbf{\hat{n}}, \Omega)  {\cal N}_\text{eq}(|\mathbf{p}|) \Delta
	\end{eqnarray*}
	In what follows we omit the dependencies on the variables for simplicity. The leading order is the second on the electron momenta and the expansion yields
    \begin{align}
    \frac{16 \pi}{3 \sigma_T}\left(1 -\frac{ \mathbf{p}}{\gamma m_ec}\cdot\mathbf{\hat{n}}\right)\frac{\id \sigma}{\id \Omega}\Delta &=\frac{x\mathbf{p}\cdot (\mathbf{\hat{n}'}- \mathbf{\hat{n}})}{m_ec}(1+\cos^2\theta) -\frac{x^2k_BT}{m_ec^2}(1-\cos\theta)(1 + \cos^2\theta)\nonumber \\ \nonumber
    +\frac{x}{(m_ec)^2}&\bigg\{(1+2\cos\theta - \cos^2\theta)(\mathbf{p}\cdot \mathbf{\hat{n}})^2 + (3 - 2\cos\theta + 5\cos^2\theta)(\mathbf{p}\cdot \mathbf{\hat{n}'})^2\;\\
    &- 4(\mathbf{p}\cdot \mathbf{\hat{n}})(\mathbf{p}\cdot \mathbf{\hat{n}'})(1+\cos^2\theta)\bigg\} \label{delta1}
    \end{align}
	where we used that $\mathbf{p}= \gamma m_e\,\mathbf{v}$.
	
	We first compute the integral over $\mathbf{p}$ in Cartesian coordinates, and then over the solid angle. Since the distribution is isotropic, the first part yields a zero contribution, \textit{i.e.},
	\begin{equation}
	\label{linearint}
	   \int \id^3\mathbf{p}\, \mathbf{p}\cdot (\mathbf{\hat{n}'}- \mathbf{\hat{n}}){\cal N}_\text{eq}(|\mathbf{p}|) =0
	\end{equation}
	The integral over the momentum in the second parcel is readily done, yielding 
	\begin{equation*}
	\int \id^3\mathbf{p}\, {\cal N}_\text{eq}(|\mathbf{p}|) = n_e
	\end{equation*}
	where we used \eqref{maxwell}.  Observe moreover
	\begin{align*}
	&(\mathbf{p}\cdot \mathbf{\hat{n}})^2 = p^2_xn^2_x + p^2_yn^2_y +  
	p^2_zn^2_z + \ \mathrm{cross \ terms \ in \ coordinates \ \ (similarly \ 
	for \ } \mathbf{\hat{n}'})\\
    &(\mathbf{p}\cdot \mathbf{\hat{n}})(\mathbf{p}\cdot \mathbf{\hat{n}'}) = p^2_xn_xn'_x + p^2_yn_yn'_y +  p^2_zn_zn'_z + \ \mathrm{similar \ to \ above}
    \end{align*}
    Cross terms in the coordinates yield zero contribution for the same reason as \eqref{linearint}. Squared terms give a similar contribution, being
 \begin{align}
	& \int \id^3\mathbf{p}\,(\mathbf{p}\cdot \mathbf{\hat{n}})^2 {\cal N}_\text{eq}(|\mathbf{p}|)= I \times {\mathbf{\hat{n}}}^2 = I\nonumber\\
	& \int \id^3\mathbf{p}\,(\mathbf{p}\cdot \mathbf{\hat{n}'})^2 {\cal N}_\text{eq}(|\mathbf{p}|)= I \times {\mathbf{\hat{n'}}}^2 = I \nonumber\\
	& \int \id^3\mathbf{p}\,(\mathbf{p}\cdot \mathbf{\hat{n}})(\mathbf{p}\cdot \mathbf{\hat{n}'}) {\cal N}_\text{eq}(|\mathbf{p}|)= I \times  \mathbf{\hat{n}}\cdot\mathbf{\hat{n}'} = I \cos \theta\nonumber\\
	&\mathrm{with} \ \ I = \int \id^3 \mathbf{p} \, p^2_x \,{\cal N}_\text{eq}(|\mathbf{p}|) = n_e m_e \, k_BT
\end{align}
    
    Using all that in $I_1(x)$ gives
    \begin{equation*}
    I_1(x)= c\frac{3\sigma_T}{16 \pi}\frac{n_e k_B\textbf{}T}{m_ec^2}\left(-x^2 + 4x\right) \left\{2\pi \int^1_{-1}d\cos\theta (1-\cos\theta)(1+\cos^2\theta)\right\}
    \end{equation*}
    and since $ \int^1_{-1}dy (1-y)(1+y^2) = 8/3$,
    we find as desired
    \begin{equation}
    I_1(x)= \frac{n_e\sigma_T c\,k_BT}{m_ec^2}\;x(4-x)
    \end{equation}
	
	\subsection{Integral $I_2$}\label{int2}
	We compute 
	\begin{eqnarray*}
	I_2(x) = c\int \id^3\mathbf{p} \id\Omega\left(1 -\frac{ \mathbf{v}}{c}\cdot\mathbf{\hat{n}}\right)\frac{\id \sigma}{\id \Omega}(\mathbf{p},\mathbf{\hat{n}}, \Omega)  {\cal N}_\text{eq}(|\mathbf{p}|) \Delta^2
	\end{eqnarray*}
The expansion up to second order in the electron momentum yields
	 \begin{align}
    \frac{16 \pi}{3 \sigma_T}\left(1 -\frac{ \mathbf{p}}{\gamma m_ec}\cdot\mathbf{\hat{n}}\right)\frac{\id \sigma}{\id \Omega}\Delta^2 =\left(\frac{x}{m_ec}\right)^2(\mathbf{p}\cdot (\mathbf{\hat{n}'} - \mathbf{\hat{n}}))^2(1+\cos^2\theta) \label{delta2}
	\end{align}	
	Using the exact same strategy as before, we have 
	\begin{align*}
    & \int \id^3\mathbf{p}\,(\mathbf{p}\cdot (\mathbf{\hat{n}'} - \mathbf{\hat{n}}))^2 {\cal N}_\text{eq}(|\mathbf{p}|)= I \times (\mathbf{\hat{n}'} - \mathbf{\hat{n}})^2 = 2I(1-\cos\theta)\\
    &\mathrm{with} \ \ I = \int \id^3\mathbf{p} \, p_x^2 \, {\cal N}_\text{eq}(|\mathbf{p}|) = n_e m_e\,k_BT
    \end{align*}
    
    This yields for $I_2(x)$
    \begin{align}
	I_2(x) &=c\frac{3\sigma_T}{16 \pi}\frac{n_ek_BT}{m_ec^2}2x^2 \left\{2\pi \int^1_{-1}d\cos\theta (1-\cos\theta)(1+\cos^2\theta)\right\}\nonumber\\
	&= \frac{n_e\sigma_T c\,k_BT}{m_ec^2}2x^2
	\end{align}
	as desired.

\subsection{The integrals and the Thomson cross section}\label{intthomson}

We demonstrate here that the integrals computed using the Thomson cross section \eqref{thomson} solely (thus neglecting the M\o ller prefactor) cannot yield the Kompaneets equation.

We calculate
\begin{eqnarray*}
	I^{\text{Th}}_\ell(x) = c\int \id^3\mathbf{p} \ \id \Omega_\text{rest}	\frac{\id \sigma^\text{Th}}{\id \Omega_\text{rest}}{\cal N}_\text{eq}(|\mathbf{p}|) \Delta^\ell
\end{eqnarray*}
with $\ell=1,2$. 
The expansion up to second order in the electron momentum yields
\begin{alignat}{2}
&\frac{16\pi}{3\sigma_T}	\frac{\id \sigma^\text{Th}}{\id \Omega_\text{rest}}{\cal N}_\text{eq}(|\mathbf{p}|) \Delta=&&\frac{x\mathbf{p}\cdot (\mathbf{\hat{n}'}- \mathbf{\hat{n}})}{m_ec}(1+\cos^2\theta) -\frac{x^2k_BT}{m_ec^2}(1-\cos\theta)(1 + \cos^2\theta)\nonumber \\ 
& &&+\frac{x}{(m_ec)^2}\bigg\{(\mathbf{p}\cdot \mathbf{\hat{n}'})^2- (\mathbf{p}\cdot \mathbf{\hat{n}})(\mathbf{p}\cdot \mathbf{\hat{n}'})\bigg\}(1+\cos^2\theta)\label{deltath1}\\
&\frac{16\pi}{3\sigma_T}	\frac{\id \sigma^\text{Th}}{\id \Omega_\text{rest}}{\cal N}_\text{eq}(|\mathbf{p}|) \Delta^2=&&\left(\frac{x}{m_ec}\right)^2(\mathbf{p}\cdot (\mathbf{\hat{n}'} - \mathbf{\hat{n}}))^2(1+\cos^2\theta) \label{deltath2}
\end{alignat}
where we have dropped the ``rest'' as subscript in the scattering angle.

We readily observe a change in the last term ($O\left(|\mathbf{p}|^2\right)$) of expression \eqref{deltath1} when comparing to \eqref{delta1}, while \eqref{deltath2} matches \eqref{delta2}. Computing the integrals in the exact same way as before leads to
\begin{align}
&I^{\text{Th}}_1(x)=	\frac{n_e\sigma_T c\,k_BT}{m_ec^2}\;x(1-x)\\
&I^{\text{Th}}_2(x)= \frac{n_e\sigma_T c\,k_BT}{m_ec^2}2x^2
\end{align}
demonstrating \eqref{firiwrong}, as desired. In particular, we observe that $I^{\text{Th}}_2(x)$ remains unchanged, \textit{i.e.}, it is insensitive to the use of the M\o ller prefactor and the frame-independent expression of Klein-Nishina cross section \eqref{klein-nishina}. Thus, the change lies on $I^{\text{Th}}_1(x)$, where we fail to account for all terms up to $O\left(|\mathbf{p}|^2\right)$, as the structure of expansion \eqref{deltath1} shows. 
	
\subsection{Nonequilibrium case}\label{intnoneq}
From now on, we denote $|\mathbf{p}|=p$. In the same spirit as before we define
\[
x = \frac{\hbar \omega}{\epsilon},\quad \Delta = \frac{\hbar (\omega'-\omega)}{\epsilon}
\]
where $\epsilon$ is the characteristic energy of the electron bath, $\epsilon = \frac{\langle p^2 \rangle }{3 m_e}$. We do the expansion of the photon distribution, which gives the same as \eqref{pdist1}. The expansion in the electron distribution, however, gives to the leading order
\begin{align}
\label{distrelect}
    &{\cal N}(p') = {\cal N}(p) - \frac{m_e\epsilon {\cal N'}(p)}{p}\Delta + \left(-\frac{(m_e\epsilon)^2 {\cal N'}(p)}{p^3} + \frac{(m_e\epsilon)^2 {\cal N''}(p)}{p^2}\right)\frac{\Delta^2}{2} \\
    &\mathrm{with} \ {\cal N'}(p)=\frac{\partial {\cal N}}{\partial p} \ \mathrm{and \ so \ on} \nonumber
\end{align}
where we use a first assumption, that the electron distribution is isotropic, ${\cal N}(\mathbf{p}) = {\cal N}(p)$,
with normalization
\begin{eqnarray}
\int \id^3 \mathbf{p}\,{\cal N}(p,t)= n_e
\end{eqnarray}

For the equilibrium case we have, of course, $\epsilon = k_BT$ and ${\cal N}(p,t)={\cal N}_\text{eq}(p)$ as in \eqref{maxwell}. One checks that in this case
\begin{align*}
    &-\frac{m_e\epsilon {\cal N}_\text{eq}'(p)}{p}={\cal N}_\text{eq}(p)\\
    &\frac{(m_e\epsilon)^2{\cal N}_\text{eq}''(p)}{p^2} - \frac{(m_e\epsilon)^2{\cal N}_\text{eq}'(p)}{p^3}={\cal N}_\text{eq}(p)
\end{align*}

Plugging into \eqref{komp-boltz} reads to the leading order
\begin{align}
    \partial_t n = \, & n'I_1\left(\Delta, {\cal N}\right) + \frac{n''}{2}I_2\left(\Delta^2, {\cal N}\right) + n(1+n)I_3\left(\Delta, \frac{{\cal N}'}{p}\right)\nonumber\\
    \label{gekomp}
    &+ n'(1+n)I_4\left(\Delta^2, \frac{{\cal N}'}{p}\right) + \frac{n(1+n)}{2}\left(I_5\left(\Delta^2, \frac{{\cal N}''}{p^2}\right) + I_6\left(\Delta^2, \frac{{\cal N}'}{p^3}\right)\right)
\end{align}
where
\begin{align*}
&I_1\left(\Delta, {\cal N}\right)= c\int \id^3\mathbf p\,\id\Omega\left(1 -\frac{ \mathbf{v}}{c}\cdot\mathbf{\hat{n}}\right) \frac{\id \sigma}{\id \Omega}\Delta {\cal N} \\
&I_2\left(\Delta^2, {\cal N}\right) = c\int \id^3\mathbf p\,\id\Omega\left(1 -\frac{ \mathbf{v}}{c}\cdot\mathbf{\hat{n}}\right) \frac{\id \sigma}{\id \Omega}\Delta^2 {\cal N}\\
&I_3\left(\Delta, \frac{{\cal N}'}{p}\right)=-c(m\epsilon)\int \id^3\mathbf p\,\id\Omega\left(1 -\frac{ \mathbf{v}}{c}\cdot\mathbf{\hat{n}}\right) \frac{\id \sigma}{\id \Omega}\Delta \frac{{\cal N}'}{p}  \\
&I_4\left(\Delta^2, \frac{{\cal N}'}{p}\right)=-c(m_e\epsilon) \int \id^3\mathbf p\,\id\Omega\left(1 -\frac{ \mathbf{v}}{c}\cdot\mathbf{\hat{n}}\right) \frac{\id \sigma}{\id \Omega}\Delta^2 \frac{{\cal N}'}{p}\\
&I_5\left(\Delta^2, \frac{{\cal N}''}{p^2}\right)=c(m_e\epsilon)^2 \int \id^3\mathbf p\,\id\Omega\left(1 -\frac{ \mathbf{v}}{c}\cdot\mathbf{\hat{n}}\right) \frac{\id \sigma}{\id \Omega}\Delta^2 \frac{{\cal N}''}{p^2}\\
&I_6\left(\Delta^2, \frac{{\cal N}'}{p^3}\right)=-c(m_e\epsilon)^2\int \id^3\mathbf p\,\id\Omega\left(1 -\frac{ \mathbf{v}}{c}\cdot\mathbf{\hat{n}}\right) \frac{\id \sigma}{\id \Omega}\Delta^2 \frac{{\cal N}'}{p^3}
\end{align*}

First and second integral are computed in the exact same way as before and gives
\begin{align}
\label{i1}
&I_1\left(\Delta, {\cal N}\right)= \frac{n_e\sigma_T c}{(m_ec)^2}\frac{\langle p^2 \rangle }{3}x(4-x)\\
\label{i2}
&I_2\left(\Delta^2, {\cal N}\right) = \frac{n_e\sigma_T c}{(m_ec)^2}\frac{\langle p^2\rangle}{3}2x^2
\end{align}

The other integrals are calculated by integration by parts and using the second assumption $\lim_{p\to \infty}p^3{\cal N}(p)=0$.\\

 To compute $I_4$ we move to spherical coordinates $\id^3\mathbf{p} = p^2\id p\, \id \Omega_p $.  Using \eqref{delta2} the integral over the electron momentum becomes
\begin{align*}
  \int (\mathbf{p}\cdot (\mathbf{\hat{n}'} - \mathbf{\hat{n}}))^2\frac{{\cal N}'}{p}\id^3 \mathbf{p}& = \int \id\Omega_p |\mathbf{\hat{n}'}-\mathbf{\hat{n}}|^2\cos^2\alpha\int_0^\infty \frac{p^2{\cal N}'}{p}p^2\id p\\
  &=\int \id\Omega_p|\mathbf{\hat{n}'}-\mathbf{\hat{n}}|^2\cos^2\alpha \left\{p^3{\cal N}\bigg{|}^\infty_0 -3\int_0^\infty p^2{\cal N}\id p\right\} \\
  &=-3\int \int_0^\infty  \id\Omega_p \id p\, |\mathbf{\hat{n}'}-\mathbf{\hat{n}}|^2\cos^2\alpha\, p^2{\cal N}\\
  &=-3\int \id^3 \, \mathbf{p}\,\, (\mathbf{\hat{p}}\cdot (\mathbf{\hat{n}'} - \mathbf{\hat{n}}))^2{\cal N}
\end{align*}
where we used \eqref{limit},  $\alpha$ for the angle between $(\mathbf{\hat{n}}-\mathbf{\hat{n}'})$ and $\mathbf{\hat{p}}=\mathbf{p}/p$. The last integral is computed in the very same fashion as we did for the equilibrium case.  We find
\begin{equation*}
    \int \id^3\, \mathbf{p}\,\, (\mathbf{\hat{p}}\cdot (\mathbf{\hat{n}'} - \mathbf{\hat{n}}))^2{\cal N} = 2n_e(1-\cos\theta)\frac{1}{3}
\end{equation*}
which after integrating the solid angle gives $I_4$,
\begin{eqnarray}
\label{i4'}
I_4\left(\Delta^2, \frac{{\cal N}'}{p}\right) =  c(m_e\epsilon)\left(\frac{x}{m_ec}\right)^2\sigma_T 2 n_e
\end{eqnarray}
Looking at expressions \eqref{i1}, \eqref{i2} and \eqref{i4'} motivates introducing 
\begin{align*}
    &\epsilon = k_BT_\text{eff} \\
    &\langle p^2\rangle = 3m_ek_BT_\text{eff}
\end{align*}
Observe that these definitions are compatible with the equilibrium case. We get now
\begin{align*}
&I_1\left(\Delta, {\cal N}\right)=  \frac{n_e\sigma_T c\,k_BT_\text{eff}}{m_ec^2}x(4-x)\\
&I_2\left(\Delta^2, {\cal N}\right) = \frac{n_e\sigma_T c\, k_B T_\text{eff}}{m_ec^2}2x^2\\
&I_4\left(\Delta^2, \frac{{\cal N}'}{p}\right) =  \frac{n_e\sigma_T c \,k_B T_\text{eff}}{m_ec^2}2x^2
\end{align*}
The other integrals are quite similar, to yield
\begin{align*}
&I_3\left(\Delta, \frac{{\cal N}'}{p}\right)=\frac{n_e\sigma_T c\,k_BT_\text{eff}}{m_ec^2}x(4-x)\\
&I_5\left(\Delta^2, \frac{{\cal N}''}{p^2}\right) + I_6\left(\Delta^2, \frac{{\cal N}'}{p^3}\right)= \frac{n_e\sigma_T c\, k_B T_\text{eff}}{m_ec^2}2x^2
\end{align*}

We substitute the values back in \eqref{gekomp} to find
	\begin{equation}
	\omega^2\frac{\partial n}{\partial t}(\omega,t)=    \frac{n_e\sigma_T c}{m_ec^2}\frac{\partial }{\partial \omega}\omega^4\left\{k_BT_\text{eff} \frac{\partial n}{\partial \omega}(t,\omega) + \hbar\left[1+n(t,\omega)\right]n(t,\omega)\right\}
	\end{equation}
which is the Kompaneets equation but with an effective (kinetic) temperature
\begin{equation*}
T_\text{eff}= \frac{\langle p^2\rangle }{3m_ek_B}
\end{equation*}

\end{document}